\documentclass[12pt, letterpaper]{iopart}

\usepackage{iopams}
\usepackage{graphicx}
\usepackage{braket}

\newcommand{\textss}[1]{\scriptsize \mbox{#1}}

\newcommand{\sign}{\mbox{sgn}}

\begin{document}

\title{Cavity-enhanced ultrafast two-dimensional spectroscopy using higher-order modes}

\author{Thomas K. Allison}

\address{Stony Brook University, Stony Brook, NY 11794-3400}
\ead{thomas.allison@stonybrook.edu}
\vspace{10pt}
\begin{indented}
\item[]August 2016
\end{indented}

\begin{abstract}
We describe methods using frequency combs and optical resonators for recording two-dimensional (2D) ultrafast spectroscopy signals with high sensitivity. By coupling multiple frequency combs to higher-order modes of one or more optical cavities, background-free, cavity-enhanced 2D spectroscopy signals are naturally generated via phase cycling. As in cavity-enhanced ultrafast transient absorption spectroscopy (CE-TAS), the signal to noise is enhanced by a factor proportional to the cavity finesse squared, so even using cavities of modest finesse, a very high sensitivity is expected, enabling ultrafast 2D spectroscopy experiments in dilute molecular beams.  
\end{abstract}

%
\vspace{1pc}
\noindent{\it Keywords}: ultrafast spectroscopy, frequency combs, cavity-enhanced spectroscopy

\vspace{1pc}
\noindent Invited submission to \emph{Emerging Leaders} issue of \jpb
%
%
%

\section{Introduction}

Spectroscopy of gas-phase atoms and molecules was essential in the development of quantum mechanics and remains essential today for fundamental studies in physics and chemistry. Particularly impactful for chemical physics have been studies on the designer species that can be produced in supersonic expansions, or molecular beams \cite{Pribble_Science1994, Lovejoy_JCP1987, Rose_JCP1989}. With molecular beam methods, one can produce cold isolated molecules, specific molecular clusters, radicals, and ions with a high degree of control \cite{Scoles_book1988}. For example, with gas-phase water clusters (H$_2$O)$_n$, one can assemble the liquid ``one-molecule at a time" \cite{Keutsch_PNAS2011}, and perform detailed systematic studies of hydrogen bond networks. Electro-spray techniques even allow the introduction of very large molecules and aggregates, with vanishing vapor pressure, into gas-phase experiments. 

While physicists have demonstrated exquisite control over gas-phase molecular samples \cite{Meerakker_NatPhys2008}, the optical spectroscopy that is performed on these systems is usually much less sophisticated than their solution phase counterparts, due to limitations imposed by the very small optical densities of dilute gases. For matter in condensed phases, coherent, all-optical, third-order spectroscopies using ultrashort pulses, such as transient absorption spectroscopy and 2D spectroscopy, have now emerged as powerful techniques for studying both structure and dynamics \cite{vanWilderen_Angewandte2015} and are widely applied to variety of problems in chemistry, physics, biology, and materials science. In contrast, most spectroscopy in molecular beams must employ so-called ``action" methods, where absorption of a photon causes a detectable change in the system, such as dissociation, ioniziation, or fluorescence. For linear spectroscopy, action methods can give a faithful representation of the absorption spectrum of the molecule \cite{Buck_ChemRev2000, Paul_JPCA1999}, but for ultrafast nonlinear spectroscopy, the comparison of gas-phase action spectra to solution phase optical spectra is highly nontrivial \cite{Kohler_AnnRevPhysChem2009, Saigusa_JPhotoChemBio2007}   and there are also gaping holes in what is measurable. For example, time-resolved photoelectron and photoion spectroscopies have been successful for studying dynamics of electronically excited molecules \cite{Stolow_ChemRev2004, Allison_JChemPhys2012}, but there is no ionization-based method for studying purely vibrational dynamics analogous to the powerful tools of ultrafast infrared spectroscopy.

Although difficult, ultrasensitive detection of optical signals from molecular beams (a.k.a. ``direct absorption") is possible, and has been used for high resolution static spectroscopy for decades \cite{Lovejoy_JCP1987, Gagliardi_Book2013}. In a recent article \cite{Reber_Optica2016} we described the extension of ultrasensitive direct absorption techniques to femtosecond time-resolved experiments, reporting cavity-enhanced optical measurements in a dilute molecular beam that are simultaneously ultrasensitive \emph{and} ultrafast. Using frequency combs and optical resonators, we performed cavity-enhanced transient absorption (CE-TAS), or simple pump-probe, measurements with a time resolution of 120 fs and a detection limit of $\Delta$OD $= 2\times 10^{-10}$, a nearly four order of magnitude improvement over the previous state of the art \cite{Schriever_RSI2008}. In this article, we describe how this technology can be applied to perform ultrasensitive 2D spectroscopy. However, instead of simply adapting standard techniques for recording 2D spectra to cavity-enhancement, we describe here a method uniquely enabled by the propagation properties of light in optical cavities. We show that using higher-order cavity modes, one can naturally record cavity-enhanced 2D signals by mixing three resonantly-enhanced frequency combs with carrier-envelope offset frequencies ($f_{\textss{CEO},1}$,  $f_{\textss{CEO},2}$, $f_{\textss{CEO},3}$) to generate a fourth resonantly enhanced frequency comb with carrier-envelope offset frequency $f^{(3)}_{\textss{CEO}} = \pm( f_{\textss{CEO},1}-f_{\textss{CEO},2}) + f_{\textss{CEO},3} $. The 2D signal is isolated from background signals via a combination of phase cycling and spatial mode-matching/phase matching. Since the three frequency combs share the same repetition rate and differ only in their carrier-envelope offset frequencies, they can be generated using just one mode-locked laser and fixed-frequency acousto-optic modulators (AOMs).

Similar to CE-TAS, the techniques described here are generally applicable to the IR, visible, and UV spectral regions, and while the primary motivation of this work is to record 2D spectroscopy signals from cold gas phase molecules and clusters, the methods also may find application in condensed phase work where higher sensitivity is needed \cite{Middleton_NatChem2012, Zanni_PNASCommentary2016, Kraack_JPCC2016}, or a robust, alignment-free instrument is desired. In section \ref{sec:phasecycling}, we describe the critical connections between phase cycling 2D spectroscopy, the nonlinear mixing of frequency combs, and the Gouy phase shifts of cavity modes. In section \ref{sec:implementation}, we discuss several possible implementations for cavity-enhancing 2D signals, and discuss their advantages and disadvantages. Section \ref{sec:conclusions} summarizes the findings of the paper and discusses bandwidth considerations and future applications.

\section{Phase cycling 2D spectroscopy from a frequency comb perspective}\label{sec:phasecycling}

For simplicity, we restrict the discussion to the case where all pump and probe pulses are linearly polarized in the same plane, but the general principles discussed here easily generalize to more complicated polarization schemes \cite{Zanni_Book2011}. Adopting the notation of Hamm and Zanni \cite{Zanni_Book2011}, the nonlinear polarization produced by a sequence of pulses $E_1$, $E_2$ and $E_3$ (or complex conjugates) arriving at the sample at $t_1$, $t_2$, and $t_3$ can be expressed as \cite{Zanni_Book2011}:
\begin{eqnarray}\label{eqn:pol}
 P^{(3)}(\vec{r},t) \propto \int_0^{\infty} dt_3 \int_0^{\infty} dt_2 \int_0^{\infty} dt_1 \sum_n R_n(t_1,t_2,t_3) \times  \nonumber \\
	 \; \; \; \; \; \; \; \; \; \; \; \; \; \; \; \; \; \; \; \; E_3(\vec{r},t-t_3)E_2(\vec{r},t-t_3-t_2)E_1(\vec{r},t-t_3-t_2-t_1)
\end{eqnarray}
where $R_n$ are the third-order system response terms that encode the molecular information of interest, and the sum is over all the double-sided Feynman diagrams that survive the rotating wave approximation, including background signals not explicitly written here such as terms proportional to $E_1 E_1^*$.  This nonlinear polarization then radiates a signal field $E^{(3)}$, which is optically detected. 

In commonly employed 2D Fourier-transform spectroscopy methods, the two pump fields $E_1$ and $E_2$ correspond to separate ultrashort pulses with adjustable, interferometrically stable, relative delay $\tau$.  The pump frequency axis of a 2D spectrum is then generated by scanning $\tau$ and performing a Fourier transform. This allows for the simultaneous combination of high pump frequency resolution and high time resolution, since only short pulses are used \cite{Shim_PCCP2009}. However, there are many terms in the sum of equation (\ref{eqn:pol}), and to collect a background-free 2D spectrum one must isolate the desired subsets of this sum. This can be done either via careful arrangement of the wave-vectors ($\vec{k}_i$) so that different terms in equation (\ref{eqn:pol}) emit phase-matched signals in different directions \cite{Asplund_PNAS2000}, or by selective modulation of the pulses combined with lock-in detection \cite{Shim_PCCP2009}. For the latter, either phase \cite{DeBoiej_ChemPhysLett1995, Tian_Science2003, Grumstrup_OptExp2007} or amplitude modulation \cite{DeFlores_OptLett2007} on one of the pump pulses may be used to separate the desired signals due to the concerted action of $E_1$ and $E_2$ from the undesired transient absorption backround signals due to each pulse acting individually.

\begin{figure}[t]
\centering
\includegraphics[width=6.0in]{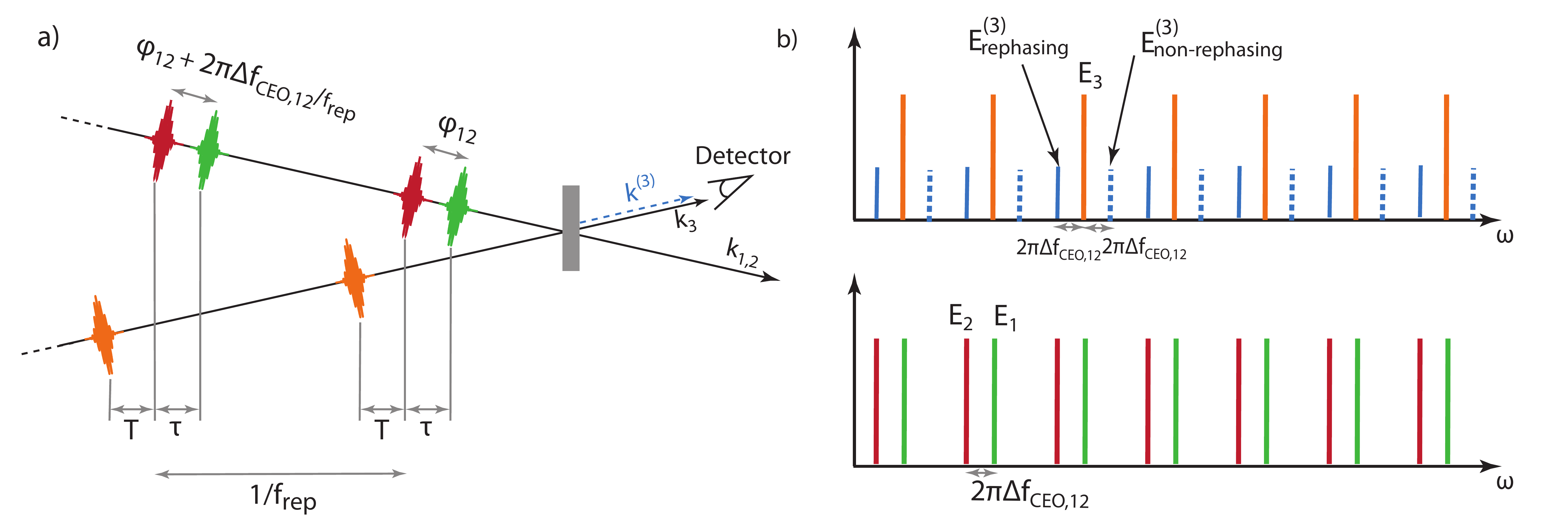}
    \caption{\small \textbf{Frequency comb perspective on phase cycling 2D spectroscopy.} Phase cycling depicted in the time a) and frequency b) domains. Two frequency combs $E_1$ and $E_2$ generate a phase cycling excitation when their carrier-envelope offset frequencies are detuned by $\Delta f_{\textss{CEO},12} \equiv f_{\textss{CEO},1} - f_{\textss{CEO},2}$. Four-wave mixing with the probe comb, $E_3$, generates new frequency combs in the $\vec{k}_3$ direction, $E^{(3)}_{\textss{rephasing}}$ and $E^{(3)}_{\textss{non-rephasing}}$. When $\Delta f_{\textss{CEO},12} \neq f_{\textss{rep}}/2$, the rephasing and non-rephasing signals show up at distinct frequencies, as shown in b), and can be separated \cite{Myers_OptExp2008, Shim_PCCP2009}. When $\Delta f_{\textss{CEO},12} = f_{\textss{rep}}/2$, such that the relative phase $ \phi_{12}$ changes by $\pi$ every $1/f{\textss{rep}}$, then the rephasing and nonrephasing signals appear at degenerate frequencies and are inseparable, giving a purely absorptive 2D signal. In both cases, a square law detector (or array) detects the heterodyne beat signal between $E_3$ and the generated third order fields at modulation frequency $\Delta f_{\textss{CEO},12}$.}
    \label{fig:phasecycling}
\end{figure}

In the phase modulation approach illustrated in figure \ref{fig:phasecycling}, commonly called ``phase cycling", the relative phase of two collinear pump pulses is varied and the spectral amplitude of the delayed probe light is detected at the modulation frequency. Mathematically, this works in the following way. The rephasing ($-\vec{k}_1 + \vec{k}_2 + \vec{k}_3$) and nonrephasing ($+\vec{k}_1 - \vec{k}_2 + \vec{k}_3$) signals are both emitted in the probe direction $\vec{k}_3$, since $\vec{k}_1 = \vec{k}_2 $. For one pulse sequence, the field emitted from the desired components of the third order polarization then depends on the carrier-envelope offset phases, $\phi_1, \phi_2, \phi_3$ of the pulses via
\begin{equation}
	E^{(3)} \propto R_{1}e^{i(-\phi_1 + \phi_2 + \phi_3)} + R_{4}e^{i(\phi_1 - \phi_2 + \phi_3)}
\end{equation}
where we are using the symbol $R_1$ here to stand in for the sum of \emph{all} the rephasing terms and $R_4$ to stand for \emph{all} the nonrephasing terms for notational simplicity, as has been adopted by other authors \cite{Zanni_Book2011}. This signal is then self-heterodyned by the probe field in a square-law detector, yielding a signal of the form \cite{Zanni_Book2011}
\begin{eqnarray}
S(\phi_1,\phi_2,\phi_3) &\propto \mathcal{R} \left[ \; E_3^* \cdot \left(R_{1}e^{i(-\phi_1 + \phi_2 + \phi_3)} + R_{4}e^{i(\phi_1 - \phi_2 + \phi_3)} \right) \;\right] \nonumber \\
 & \propto \mathcal{R} \left[ \; \left(R_{1}e^{-i\phi_{12}} + R_{4}e^{i\phi_{12}} \right) \; \right]
\end{eqnarray}
where $\mathcal{R}$ denotes the real part and $\phi_{12} \equiv \phi_1 - \phi_2$. By constructing linear combinations of signals with different phases $\phi_{12}$ one can recover the rephasing and nonrephasing components of the 2D spectrum or any desired combination \cite{Shim_PCCP2009, Zanni_Book2011}. For example, to record purely absorptive 2D spectra, one commonly records signals with phase differences $\phi_{12}  = 0$ and $\pi$:
\begin{equation}\label{eqn:piphase}
	S_{\textss{absorptive 2D}} = S(\phi_{12} = 0) - S(\phi_{12}= \pi)
\end{equation}
The desired 2D signals add in this construction, while unwanted background signals due to the action of one pump pulse alone are subtracted away. One can also use phase differences other than 0 and $\pi$ to recover the rephasing and non-rephasing signals separately \cite{Zanni_Book2011, Myers_OptExp2008}. We will return to this point below.

Since the excitations probed in 2D spectroscopy typically decohere on picosecond time scales, the mutual coherence of successive pulse sequences at repetition rate $f_{\textss{rep}}$, usually separated by milliseconds, is of no consequence. The sample has no coherent memory of the last pulse sequence, so it does not matter whether there is a definite phase relationship between $E_1$ pulses $n$ and $n+1$ or not. Put another way, it does not matter if $E_1$ is a phase-coherent frequency comb. The only coherence that matters is that for every pulse sequence, there is a definite phase relationship between the pulses $E_1$ and $E_2$, separated on the ultrafast time scale. However, to understand how 2D spectroscopy signals can be cavity-enhanced, it is instructive to consider the standard phase-cycling experiment described by equation (\ref{eqn:piphase}) in the case where pulses separated by $1/f_{\textss{rep}}$ \emph{are} coherent, and $E_1$, $E_2$ and $E_3$ do constitute frequency combs. Consider the case where $\phi_{12}$ is incremented by the phase shift $\Delta\phi_{12}$ every sequence of laser pulses. Since the carrier envelope offset frequency of a frequency comb, $f_{\textss{CEO}}$, is simply given by 
\begin{equation}
	f_{\textss{CEO}} =  f_{\textss{rep}}\frac{\Delta \phi_{\textss{CE}}}{2\pi}
\end{equation}
where $\Delta \phi_{\textss{CE}}$ is the pulse-to-pulse carrier envelope phase shift and $f_{\textss{rep}}$ is the comb's repetition rate, the combs $E_1$ and $E_2$ share the same repetition rate but differ in their carrier-envelope offset frequency by $f_{\textss{CEO},1}-f_{\textss{CEO},2} =\Delta \phi_{12}f_{\textss{rep}}/2\pi$, as illustrated in figure \ref{fig:phasecycling}b). These two pump combs then mix with the probe comb via the third-order response, with offset frequency $f_{\textss{CEO},3}$ to produce new combs with offset frequencies 
\begin{equation}\label{eqn:f0cond}
	f_{\textss{CEO}}^{(3)} = \mp f_{\textss{CEO},1} \pm f_{\textss{CEO},2} + f_{\textss{CEO},3} = \mp f_{\textss{rep}}\frac{\Delta \phi_{\textss{12}}}{2\pi} + f_{\textss{CEO,3}}
\end{equation}
where the upper sign corresponds to the new comb generated via the rephasing components of the third order response and the lower sign to the nopnrephasing components. This is illustrated in the time and frequency domains in figure \ref{fig:phasecycling}. The pulses of the probe comb $E_3$ and the generated combs $E^{(3)}$ are coincident in time, and thus give rise to heterodyne beat signals at the differences between their offset frequencies. For the common case where $\Delta \phi_{12} = \pi$, the intensity at the square-law detector is thus modulated at $f_{\textss{rep}}/2$ and the absorptive 2D spectroscopy signal is isolated from the background DC signal via lock-in detection (i.e. differencing) at $f_{\textss{rep}}/2$. Just as the rephasing and nonrephasing signals are added in the conventional phase cycling method with $\Delta \phi_{12} = \pi$, one can see from figure \ref{fig:phasecycling}b) that in this case where $|f_{\textss{CEO},1}-f_{\textss{CEO},2}| = f_{\textss{rep}}/2$ the generated rephasing and nonrephasing signal combs are degenerate, and the signals are coherently combined.  To separate the rephasing and nonrephasing signals, one phase cycles with $\Delta \phi_{12} \neq \pi$, or offset frequency difference other than $f_{\textss{rep}}/2$, in which case the combs generated via rephasing and norephasing contributions to the third order response appear at nondegenerate frequencies, and can be separated \cite{Zanni_Book2011,Myers_OptExp2008}.

Once this connection between phase cycling and wave-mixing of frequency combs is understood, it becomes clear how to cavity-enhance the phase-cycling 2D signal: one tunes the modes of one or more optical cavities such that all four frequency combs, both the three combs provided by the spectroscopist ($E_1$, $E_2$, $E_3$) and the generated $E^{(3)}$, are all resonant with modes of optical cavities. As in CE-TAS, the sample has no memory of prior pump and probe pulses, but the cavity does. Each field is enhanced by a factor proportional to the square root of the cavity finesse\footnote{The field generated in the cavity, $E^{(3)}$ is actually enhanced $\propto \mathcal{F}$ inside the cavity, but then must be reduced $\propto \sqrt{\mathcal{F}}$ for detection outside the cavity, so that the overall field enhancement for the generated $E^{(3)}$ field (not counting the power enhancements of the driving fields) scales as $\sqrt{\mathcal{F}}$}, $\sqrt{\mathcal{F}}$, so in the limit that the molecular excitation is not saturated the attainable signal to noise scales as $\mathcal{F}^2$, as in CE-TAS \cite{Reber_Optica2016}. Thus, even for cavities of modest finesse, very large signal enhancements are possible. There are, in principle, several ways one could achieve this resonance condition. In this article we focus on using higher-order cavity modes, which allows the FSR of the cavities employed to remain matched to the frequency comb, providing the optimum enhancement of the intracavity peak power.\footnote{One could also, in principle, use an overly long cavity such that $f_{\textss{rep}}$ of the comb is an integer multiple of the cavity FSR. This would provide extra TEM$_{00}$ resonances for coupling multiple, $f_{\textss{CEO}}$-shifted combs to the same cavity. However, this method would suffer several drawbacks. First, since there are multiple pulse sequences per round trip circulating in the cavity, the peak power of both the pump and probe pulses is less, lowering the nonlinear signal size. Second, the cavity linewidth is narrower, increasing the technical difficulty without increasing the signal size.} 

In section \ref{sec:implementation}, we describe several possible physical implementations, but here we first discuss the basic premise of the idea. Phase cycling using higher-order modes is motivated by the mode structure of optical resonators. In an optical cavity, light pulses in different spatial modes acquire a round trip differential phase shift due to the dependence of the round trip Gouy phase on the Hermite-Gaussian mode order. In general, if $E_1$ is in the TEM$_{l_1 m_1}$ mode and $E_2$ is in the TEM$_{l_2 m_2}$ mode, each round trip they acquire a phase shift
\begin{equation}\label{eqn:psi}
	\left. \Delta\phi_{12} \right|_{\textss{round trip}} = (l_1-l_2)\psi_{\textss{tan}} + (m_1-m_2)\psi_{\textss{sag}}
\end{equation}   	    
with the Gouy phase shifts $\psi_{\textss{tan}}$ and $\psi_{\textss{sag}}$ solely determined by the geometry of the cavity, related to the components of the $ABCD$ matrices via $\psi = \sign(B) \cos^{-1} \left[(A+D)/2 \right]$, with separate $ABCD$ matrices for the sagittal and tangential planes, respectively. These phase shifts are tunable. For example, for a simple resonator with two concave mirrors of equal curvature, $\psi = \psi_{\textss{tan}} = \psi_{\textss{sag}}$ is continuously tunable from near $0^{+}$ to $+\pi$ (near planar $\rightarrow$ confocal) and $0^{-}$ to $-\pi$ (near concentric $\rightarrow$ confocal). Since the Gouy phase shift depends only on the cavity geometry and is independent of wavelength, it corresponds to a pure carrier-envelope offset frequency shift, viz.
\begin{equation}\label{eqn:psi}
	f_{\textss{CEO},1}-f_{\textss{CEO},2} = \frac{f{\textss{rep}} }{2\pi} \left. \Delta\phi_{12} \right|_{\textss{round trip}}
\end{equation}   
Thus, by coupling combs to the higher-order modes of an optical cavity they naturally phase cycle, generating new combs which can also be made resonant. As we discuss in the next section, mode-matching also provides spatial isolation of the signal analogous to non-collinear phase matching in conventional 2D spectrometers.

\section{Implementations}\label{sec:implementation} 

As in conventional 2D spectroscopy setups using mJ-pulsed lasers, there are many conceivable physical implementations of the resonantly enhanced phase-cycling scenario discussed above. In general, since $E_1$, $E_2$, and $E_3$ share the same repetition rate and only differ by their carrier envelope offset frequency, they can be generated from one mode-locked laser simply by diffraction from fixed-frequency AOMs, and one does \emph{not} need three separate frequency comb lasers. The optimum choice of cavity geometry and mode selection depends on several factors, including system complexity, signal enhancement factor, signal specificity, ease of alignment, attainable sample length, and signal readout. Design decisions will thus likely be driven by the demands of a particular measurement. In this section, we discuss several possible implementations and their relative strengths and weaknesses.

We restrict the discussion to bow-tie ring cavities for the reason that they allow independent control of the overall cavity length and the focus size. This allows one to separately control the peak intensity at the sample and the repetition rate of the system. Ring cavities also allow for the easy introduction of counter-propagating reference beams for common-mode noise subtraction, as has been critical for the success of CE-TAS \cite{Reber_Optica2016}. For a bow-tie ring cavity, the sign of the $B$ component of the $ABCD$ matrix is always negative, such that phase shifts $\psi_{\textss{tan}}$ and $\psi_{\textss{sag}}$ are restricted to the range between 0 and $-\pi$. Figures \ref{fig:onecav} and \ref{fig:twocav} show implementations of cavity-enhanced 2D spectroscopy using one and two ring cavities, respectively. Both generate signals that are ``background-free" in the sense that the signal field is generated in an unoccupied cavity mode. Using one cavity makes the optical alignment and stabilization of the system very simple, and also permits the use of an extended slit jet expansion for an increased column density of molecules \cite{Busarow_JChemPhys1988}, but requires separation of the weak signal field from the intense collinear pump and probe fields. Using two cavities makes the alignment and stabilization more complicated but it is easier to isolate the desired 2D signal. We discuss these subtleties in more detail in the sections below.

\subsection{One-cavity schemes}

\begin{figure}[t]

\centering
\includegraphics[width=3.25in]{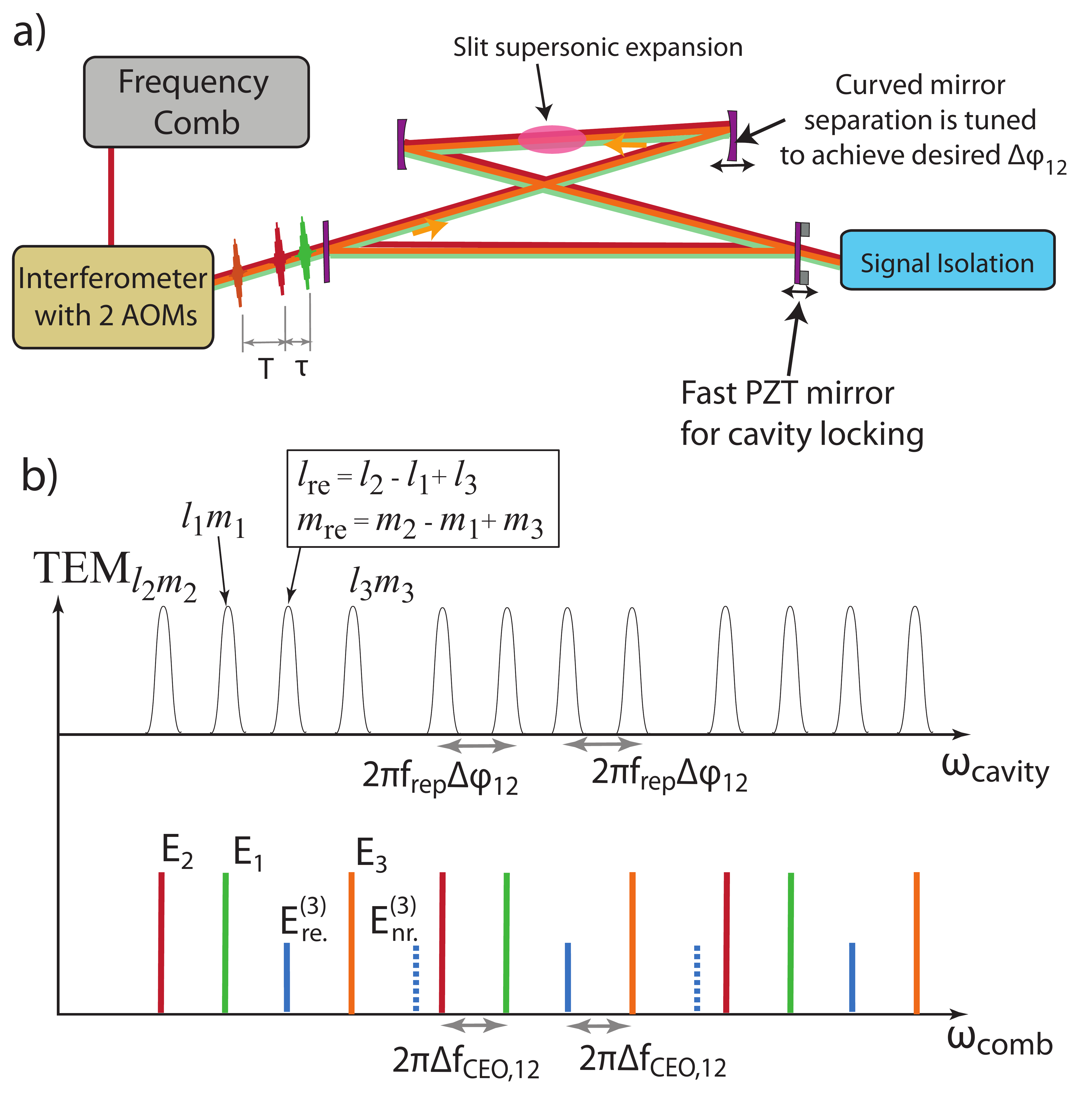}
    \caption{\small \textbf{Cavity-enhanced 2D spectroscopy using one cavity.} a) A frequency-comb and an interferometer with two AOMs are used to generated three frequency combs with distinct carrier-envelope offset frequencies that are then resonantly enhanced in a passive optical cavity, generating a resonantly-enhanced 2D spectroscopy signal if the resonance conditions illustrated in b) are met. The one-cavity scheme simplifies the alignment and laser/cavity stabilization and also allows the use of an extended sample (slit expansion), but has the drawback that the cavity's transmitted light must be well resolved to separate the weak signal from the strong pump and probe pulses. By choosing the symmetry of the excited cavity modes and tuning the resonance frequencies, the spectroscopist can select what signals are resonantly enhanced and suppress background. Shown is a case where the rephasing signal is resonantly enhanced but	 the non-rephasing signal is not. In this case, the non-rephasing signal can still be recorded by reversing the time-ordering of $E_1$ and $E_2$.}
\label{fig:onecav}
\end{figure}

In the one-cavity scheme illustrated in figure \ref{fig:onecav}, three collinear frequency combs with different f$_{\textss{CEO}}$'s are coupled to three different Hermite-Gaussian spatial modes of a ring cavity with normalized field amplitudes described mathematically at the beam waist via \cite{Siegman_book1986}:

\begin{equation}
	\fl u_{lm}(x,y) = \left( \frac{2}{\pi} \right)^{1/2} \sqrt{\frac{1}{2^{(l+m)} \; w_{0x}l! \;w_{0y} m! }} H_l \left( \frac{\sqrt{2}x}{w_{0x}} \right) H_m \left( \frac{\sqrt{2}y}{w_{0y}} \right)e^{-x^2/w_{0x}^2}e^{-y^2/w_{0y}^2}
\end{equation}
where $l$ and $m$ are the mode orders in the tangential (x) and sagittal (y) planes, respectively, $H_l$ is the $l^{th}$ order Hermite polynomial, and $w_{0x}$ and $w_{0y}$ are the $1/e^2$ intensity radii of the fundamental TEM$_{00}$ mode in the $x$ and $y$ directions. In a ring cavity with spherical mirrors, astigmatism causes $w_{0x} \neq w_{0y}$, which breaks the degeneracy between horizontal and vertical modes via their different round-trip Gouy phase shifts, described by equation (\ref{eqn:psi}). To resonantly enhance a desired 2D signal, the generated comb must be resonant with one or more of the cavity's transverse modes. Analogous to equation \ref{eqn:f0cond}, mathematically, this means that there exists at least one set of integers $l_t$ and $m_t$ for the target mode that satisfy
\begin{equation}
	l_t \psi_{\textss{tan}} + m_t \psi_{\textss{sag}} = (\mp l_1 \pm l_2 + l_3)\psi_{\textss{tan}} + (\mp m_1 \pm m_2 + m_3)\psi_{\textss{sag}}
\end{equation} 
where the upper sign corresponds to the rephasing signal, and the lower sign corresponds to the non-rephasing signal. This can be satisified  in simple fashion via $l_t =\mp l_1 \pm l_2 + l_3$ and $m_t = \mp m_1 \pm m_2 + m_3$, as shown in figure \ref{fig:onecav}, but can also be satisfied in other ways, particularly when either $2\pi/\psi_{tan}$ or $2\pi/\psi_{sag}$ are integers and several modes are degenerate. For example, with $E_1$ in the TEM$_{10} $ mode, $E_2$ in the TEM$_{11}$ mode, $E_{3}$ in the TEM$_{00}$, the rephasing signal is clearly resonant with the TEM$_{01}$ mode since $-l_1 + l_2 + l_3 = 0$ and  $-m_1 + m_2 + m_3 = 1$. In contrast, the simple arithmetic for the non-rephasing signal gives $(l_t,m_t) = (0,-1)$, and there is no TEM$_{0,-1}$ mode, and it appears that this signal is not resonant, as illustrated in figure \ref{fig:onecav}. However, if for example $\psi_{\textss{sag}} = -\pi/2$, the the non-rephasing signal is resonant with the TEM$_{03}$ mode, which also has the appropriate even-x, odd-y symmetry to accept the signal.

Only the spatial component of the generated field that is mode-matched to the target cavity mode will be resonantly enhanced. The spatial overlap factor $\braket{u_t | u^{(3)}}$ between the generated $E^{(3)}$ comb, with normalized spatial mode amplitude $u^{(3)}$, and the target resonant TEM$_{l_tm_t}$ mode, with normalized spatial amplitude $u_t$, is given by
\begin{equation}\label{eqn:modematch}
	\braket{u_t | u^{(3)}} = \frac{\int dx \int dy \; u_t^* u_{l_1m_1}u_{l_2m_2}u_{l_3m_3}}{\int dx \int dy \; u_{l_1m_1}^*u_{l_2m_2}^*u_{l_3m_3}^* u_{l_1m_1}u_{l_2m_2}u_{l_3m_3} }
\end{equation}
Now the generated signal field $E^{(3)}$ is enhanced by a total factor proportional to $|\braket{u_t | u^{(3)}}|(\mathcal{F}/\pi)^2$. Indeed, this would imply that if one detects the intensity of the generated light on its own (homodyne detection), for example with a VIPA spectrometer \cite{Shirasaki_OptLett1996} or spatial-mode division multiplexing \cite{Ryf_JLightwaveTech2012}, then the signal in fact scales as $I^{(3)} \propto |E^{(3)}|^2 \propto |\braket{u_t | u^{(3)}}|^2(\mathcal{F}/\pi)^4$. Each of the three input beams in the driven four-wave mixing process \cite{Shen_book1984} has an intensity enhancement of $\mathcal{F}/\pi$ and the cavity also provides an additional enhancement of $|\braket{u_t | u^{(3)}}|^2\mathcal{F}/\pi$ for the intensity of the generated light, giving an overall scaling of $|\braket{u_t | u^{(3)}}|^2(\mathcal{F}/\pi)^4$. However due to the expected small absolute size of the signal, it is still likely advantageous to employ heterodyne detection of the generated field. Indeed, conventional background-free 2D spectroscopy, isolated by phase matching, is still in general less sensitive than heterodyne detected signals recorded in a pump-probe geometry \cite{Bizimana_JChemPhys2015, Arpin_JPhysChemB2015}. In both cases, the fundamental shot-noise limit on the signal to noise scales only as $|\braket{u_t | u^{(3)}}|(\mathcal{F}/\pi)^2$, since in the heterodyne case the noise level is determined by the noise of the local local oscillator, but in the intensity (homodyne) measurement, the noise scales as $\sqrt{I^{(3)}}$. Also note that although the $E^{(3)}$ field enhancement is only reduced by one power of the mode-matching factor $|\braket{u_t | u^{(3)}}|$, an additional mode-matching factor less than unity may be encountered in heterodyne detection. For example, if the heterodyne beat between $E^{(3)}$ and $E_3$ fields is detected by simply recording the amplitude modulation on the probe beam in the two-cavity scheme (figure \ref{fig:onecav}), the orthogonality of the Hermite-Gaussian modes requires sampling of less than the whole beam to recover a non-zero beat signal. 

Equation (\ref{eqn:modematch}) also provides an opportunity to understand the physical origin of the signal from the perspective of the probe pulse absorption and diffraction. In conventional third-order spectroscopy setups using free-space non-collinear beams, one can think of the pump pulse(s) generating a spatially dependent excitation pattern that the probe light can be diffracted from. When the pump pulse(s) overfill the volume of the sample probed by the probe beam in a pump-probe geometry, only the probe absorption is modulated, its spatial mode unchanged. However, if the coherent excitation of the medium by the pump pulse(s) is not spatially uniform, a transient excitation grating is created which can diffract the probe beam into different directions, producing a background-free signal. The mixing of higher-order cavity modes can be viewed in this way. The two pump fields $E_1$ and $E_2$ act in concert to produce a spatially non-uniform excitation, which causes the probe light to diffract into the higher order modes of the cavity. For resonance, the spatial pattern is modulated (via $\Delta f_{\textss{CEO,12}}$) such that the diffracted probe pulses from successive round trips interfere constructively, and the diffracted probe beam is then resonant with one of the modes of the cavity. Indeed, one can see from equation (\ref{eqn:modematch}) that spatial inhomogeneity of the pump fields is crucial for generating a non-zero signal. If the pump modes overfill the probe volume such that $u_{l_1m_1}u_{l_2m_2} \rightarrow const.$ in equation (\ref{eqn:modematch}), then $|\braket{u_t | u^{(3)}}|$ is identically zero due to the orthogonality of the Hermite-Gaussian modes!\footnote{For our previous CE-TAS demonstration \cite{Reber_Optica2016}, the modulation frequency was much less than the cavity linewidth, and thus the target mode is the same as the probe mode ($u_t = u_3$) and then  $|\braket{u_t | u^{(3)}}| \rightarrow 1$ in the limit of pump overfill.} 

The spatial-mode selectivity of the cavity via equation (\ref{eqn:modematch}) is analogous to phase matching in conventional 2D spectroscopy setups. Just as one isolates a desired signal in a boxcar geometry by detecting in a certain direction, in CE-2D spectroscopy using higher-order modes one can isolate a desired signal by detecting in a certain spatial mode. The generation and resonant enhancement of CE-2D signals using higher order cavity modes can thus be viewed as selecting a desired third-order response signal through a combination of \emph{both} phase cycling and spatial discrimination/phase matching. This combination can make CE-2D spectroscopy highly selective, even in the completely collinear geometry of fig. \ref{fig:onecav}a), since both the cavity and the detection methods \cite{Coddington_Optica2016, Maslowski_PRA2016,Ryf_JLightwaveTech2012} can discriminate against undesired signals. As an example, lets again consider the simple case where $E_1$, $E_2$, and $E_3$ are coupled into the TEM$_{10}$, TEM$_{11}$, and TEM$_{00}$ modes of one optical cavity. The rephasing signal is resonantly enhanced in the TEM$_{01}$ mode with a mode-matching factor $|\braket{ u_t | u^{(3)} } |= 0.65$. Without mode degeneracy, the non-rephasing signal is not resonantly enhanced, and would instead be recorded by reversing the time-ordering of $E_1$ and $E_2$ \cite{Zanni_Book2011}. Many undesired signals, although emitted collinearly, are suppressed from the target mode via a combination of the spatial and frequency discrimination. For example the transient absorption signals $\propto |E_1|^2 E_3$ and $\propto |E_2|^2E_3$ are enhanced in the TEM$_{00}$ mode occupied by $E_3$ but are generated with both the wrong frequency ($f^{(3)}_{\textss{CEO}} =  f_{\textss{CEO,3}} \neq f_{\textss{CEO,3}} - \frac{f{\textss{rep}}}{2\pi} \left. \Delta\phi_{12} \right|_{\textss{round trip}}$) and the wrong spatial symmetry ($\braket{ u_{01} | u^{(3)} } = 0$) to appear in the target TEM$_{01}$. Similarly, 2 quantum signals $\propto E_1 E_2 E_3^*$ are weakly resonant with the TEM$_{21}$  mode ($|\braket{ u_{21} | u^{(3)} } |= 0.05$) but are suppressed from the target TEM$_{01}$ mode by frequency discrimination. Some fifth-order signals and cascaded third-order signals do satisfy the necessary resonance and symmetry requirements to be resonantly enhanced in the target mode, but can be distinguished via power and sample density dependence of the signal, as in conventional 2D spectroscopy. 

Using a cavity where some of the modes are degenerate provides additional opportunities \cite{Weitenberg_OptExp2011}. This can be done by tuning the curved mirror separation, $\delta$, such that either $2\pi/\psi_{tan}$ or $2\pi/\psi_{sag}$ (or both) are integers. For cavities with long focal length mirrors, as we employed in CE-TAS \cite{Reber_Optica2016}, this does not require particularly precise control of the curved mirror separation. For example, for a four-mirror bow-tie cavity with FSR = 87 MHz, 75 cm mirror radius of curvature mirrors, and the curved mirror seperation $\delta = 90.9$ cm (the center of the stability region), $\psi_x = -\pi/2$, and $d\psi_x/d\delta$ is only 0.06 rad/cm. Since a $2\pi$ intracavity phase shift corresponds to one cavity FSR, this corresponds to a frequency shift of only 0.01 FSR/cm. The cavity linewidth is the FSR divided by the finesse, so for a cavity finesse of 1000, a frequency shift of one cavity linewidth corresponds to a large change in $\delta$ of 1 mm. For achieving degeneracy of 5 higher-order transvers modes, separated by $\Delta l = 4$, all within 1/10 of a cavity linewidth thus only requires control of $\delta$ to the length scale of 2 $\mu$m, which can likely even be achieved passively with careful design. To put this in perspective, to lock the cavity to the comb, the overall length of the cavity is already being actively stabilized to much better than $\lambda/\mathcal{F} <$ 1 nm.

With degenerate modes, one can record purely absorptive 2D spectroscopy signals using $|f_{\textss{CEO,1}} - f_{\textss{CEO,2}}| = f_{\textss{rep}}/2$ and have the rephasing and non-rephasing signals constructively interfere in the same set of cavity modes. Mode degeneracy allows for somewhat improved mode-matching, since both the driving fields and the generated field can exist in a superposition of degenerate spatial modes. For example consider the case where $\psi_x = -\pi/2$, and the spectroscopist couples $E_1$ to a superposition of the first five degenerate TEM$_{3+4n,0}$ and $E_2$ to a superposition of the first five degenerate TEM$_{1+4n,0}$, so that $\Delta \phi_{12}|_{\textss{round trip}} = - \pi$ and $|f_{\textss{CEO,1}} - f_{\textss{CEO,2}}| = f_{\textss{rep}}/2$. With $E_3$ coupled to a superposition of the first five degenerate TEM$_{0+4n,0}$ modes, we find via rough numerical optimization of equation (\ref{eqn:modematch}) that mode-matching factors $|\braket{ u_t | u^{(3)} }|^2 > 0.7$ are attainable, with $u_t$ a superposition of the first five TEM$_{2+4n,0}$ modes. \footnote{Note that here we report $|\braket{ u_t | u^{(3)} }|^2$ for the total power enhancement instead of $|\braket{ u_t | u^{(3)} }|$ for the field enhancement because in the superposition state the field enhancement is spatially dependent (in $x,y$, and $z$) and is somewhat meaningless, whereas in the single-mode situation it is reasonably straightforward in certain cases to achieve mode-matched heterodyne detection, measuring the field $\propto |\braket{ u_t | u^{(3)} }|$ .}

The above cases illustrate that selectively cavity-enhancing the desired 2D spectroscopy signals, either purely absorptive signals or separate components, can be done well with one cavity in a collinear geometry. The main challenge of using one cavity will be the separation of the miniscule 2D spectrosocpy signal from the much more intense intense input beams. $E_3$ is distinct from the signal field via spatial mode and frequency. $E_1$ and $E_2$ are distinct via spatial mode, frequency, and also in the time domain for waiting times $T > 0$.  Thus, in principle, the signal isolation problem has already been solved by the practitioners of mode-division multiplexing \cite{Ryf_JLightwaveTech2012}, in which signals in different spatial modes are de-multiplexed at the end of multi-mode fiber, and direct frequency comb spectroscopy (DFCS) where individual comb-teeth are resolved \cite{Adler_AnnRevChem2010, Coddington_Optica2016, Maslowski_PRA2016}. However, some simple estimates indicate that a very high degree of isolation will be necessary. In our demonstration of CE-TAS in a dilute molecular iodine sample, the power in the resonantly enhanced, nonlinearly (third-order) generated AM sidebands on the probe light outside the cavity was less than 100 fW. Lock-in detection at the heterodyne beat frequency between the AM sidebands and the $\sim 1$ mW probe beam provides passive amplification of the signal to detectable levels in a balanced detection scheme. But in the one-cavity scheme, this 100 fW should be compared to the $\sim 1 $ W of incident pump light used to generate it. Clearly, in practice very strong suppression of undesired signals and noise is necessary.

\begin{figure}[t]
\centering
\includegraphics[width=3.25in]{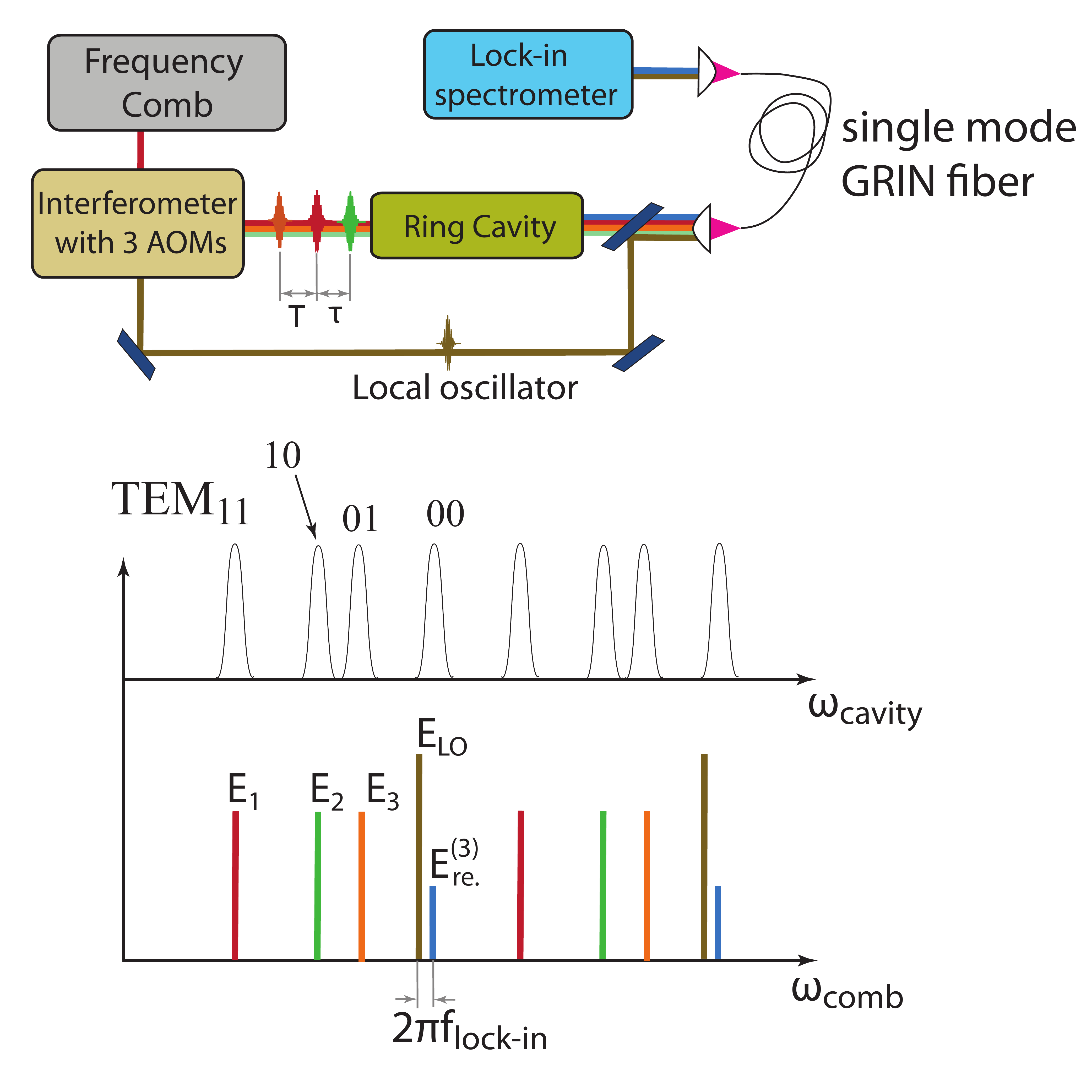}
    \caption{\small \textbf{Isolation of the 2D signal.} The generated frequency comb $E^{(3)}$ is different from the three driving fields in time, spatial mode, and frequency. The proposed scheme shown here uses all three to isolate the 2D signal from background. First, the TEM$_{00}$ beam from the cavity is selected by coupling the cavity output to a single mode graded index fiber. Second, heterodyne detecting the signal with a local oscillator (LO) frequency comb and a lock-in spectrometer discriminates against the collinear combs $E_1$, $E_2$, and $E_3$ in the frequency domain, since their heterodyne signals with the LO comb appear at the wrong frequencies. Third, background signals due to $E_1$ and $E_2$ are further suppressed by adjusting the delay of the LO pulses such that they coincide temporally with only the $E^{(3)}$ and $E_3$ combs.}
    
\label{fig:onecavdet}
\end{figure}

As in DFCS, there are likely several ways to isolate the desired signal from the background, and the optimum configuration likely hinges on experimental details not yet anticipated, and will require further study. However, one has a major advantage here over DFCS in that high spectral resolution of the signal comb is not necessary, as the features probed in ultrafast nonlinear spectroscopy are generally broad due to fast decoherence and population relaxation. In figure \ref{fig:onecavdet}, we propose a simple signal isolation scheme based on a combination of spatial-mode discrimination and frequency-discrimination. The input fields $E_1$, $E_2$, and $E_3$ are coupled to the TEM$_{11}$, TEM$_{10}$, and the TEM$_{01}$ modes, respectively, generating a rephasing signal in the lowest-order TEM$_{00}$ mode with $|\braket{ u_t | u^{(3)} } |= 0.65$. Light from the cavity is then launched in a single-mode graded index fiber to remove the $E_1$, $E_2$, and $E_3$ fields from the beam. Graded index fiber is used because the modes of graded index fiber are Hermite-Gaussians \cite{Marcuse_JOSA1978}, and thus form the same orthonormal basis as that of the cavity, providing (theoretically) perfect rejection of the unwanted higher order modes. Also coupled to the fiber is an intense local oscillator frequency comb with $f_{\textss{CEO},LO}$ near $f_{CEO}^{(3)}$. After exiting the fiber, the modulation on the light at frequency $f_{\textss{CEO},LO}$ - $f_{\textss{CEO}}^{(3)}$ is detected using a lock-in spectrometer, providing discrimination in the frequency domain. A lock-in spectrometer can be achieved trivially using a scanning grating monochromator or Fourier-transform spectrometer with a fast single-element detector, or nontrivially using specialized lock-in array detectors \cite{Sinclair_PRL2011,Light_SPIE2010} or time-stretch dispersive Fourier transform techniques \cite{Herink_NatPhot2016,Kelkar_ElectronicsLetters1999} for parallel detection. For waiting times $T>0$, this scheme additionally discriminates in the time domain due to the fact that the local-oscillator does not overlap temporally with $E_1$ and $E_2$.

\subsection{Two-Cavity Schemes}

\begin{figure}[t]
\centering
\includegraphics[width=6.5in]{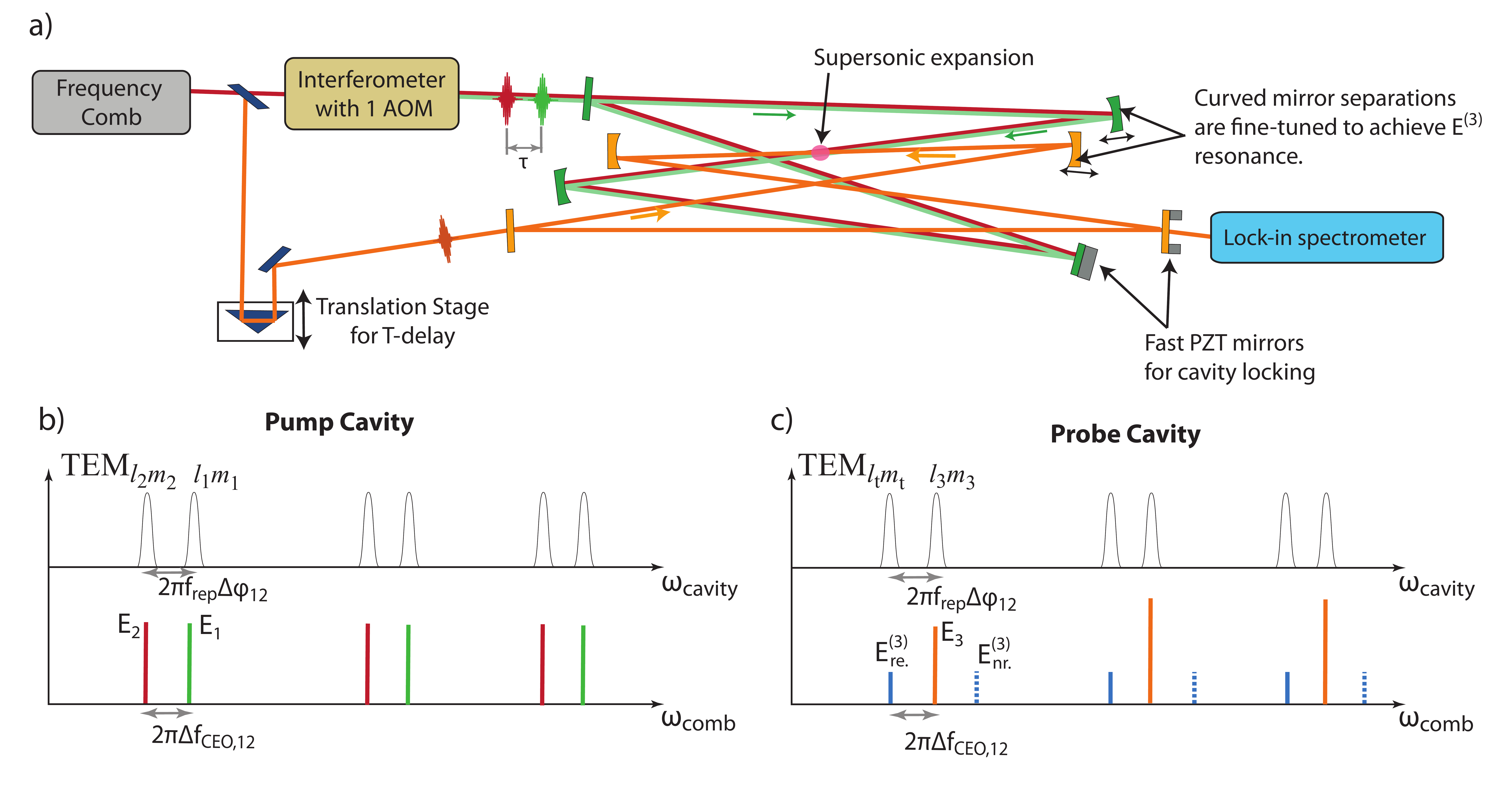}
    \caption{\small \textbf{Cavity-enhanced 2D spectroscopy in a pump-probe geometry using two cavities.} a) Light from a frequency comb source is split into pump and probe beams. An interferometer with a fixed-frequency acousto-optic modulator (AOM) in one arm generates a pair of pump frequency combs shifted by the AOM drive frequency and variable delay. A background-free 2D spectroscopy signal is passively amplified in the probe cavity if the necessary resonance conditions illustrated in b) and c) are met by appropriate tuning of the frequency comb, AOM frequency, and cavity-mirror separations. The signal is recorded via lock-in detection of the heterodyne beating between the generated $E^{(3)}$ comb and the probe light. }
\label{fig:twocav}
\end{figure}

Much of the discussion for one cavity schemes carries over to two-cavity schemes, illustrated in figure \ref{fig:twocav}. Good mode-matching and background suppression can be achieved through appropriate mode selection, and degenerate modes can be used to record purely absorptive 2D spectroscopy signals. The pump-probe geometry also makes it much easier to use very different frequencies for pump and probe, such as employed in 2D electronic-vibrational spectroscopy \cite{Oliver_PNAS2014}. But the main advantage of using two cavities is that the non-collinear pump and probe allows for much easier isolation of the desired 2D signal via simple lock-in detection at the phase cycling frequency $|f_{\textss{CEO,1}} - f_{\textss{CEO,2}}|$. This is similar to the amplitude modulation frequency scheme used in our previous CE-TAS demonstration and thus one would expect similar signal to noise considerations, except that now the phase cycling scheme has a distinct advantage: the modulation frequency is now much larger than the cavity linewidth. This means that the signal appears at a frequency where there is naturally greatly reduced intensity noise on the transmitted light because the cavity low-pass filters the laser noise at Fourier frequencies much larger than the cavity linewidth \cite{Nagourney_Book2010, Lawrence_JOSAB1999}!\footnote{This can also be exploited in the heterodyne detection scheme with a separate local-oscillator of figure \ref{fig:onecavdet}}.

The price one pays for simplified detection of the 2D signal is complexity in the optical layout and its alignment. With two cavities, their foci must be overlapped in the molecular beam, and the required finite crossing angle means a smaller overlap volume can be achieved in the molecular beam sample. There are also now two cavities that require stabilization and tuning of $\psi_{\textss{sag}}$ and $\psi_{\textss{tan}}$, although if the $f_{\textss{CEO}}$ frequencies of the two cavities can be matched, as in \cite{Reber_Optica2016}, then in principle one can use fewer AOMs.

\subsection{Interferometers and mode conversion}

For completeness, we briefly describe some possibilities for generating multiple collinear frequency combs with different and tunable $f_{\textss{CEO}}$'s and adjustable delays and in different spatial modes for efficient mode-matching to the cavity. This can be done in reasonably straightforward fashion by incorporating AOMs and phase/amplitude masks \cite{Ryf_JLightwaveTech2012} into a stabilized ``pulse stacker" \cite{Cryan_PRA2009, Kienel_OptLett2014, Siders_ApplOpt1998,DeBoiej_ChemPhysLett1995}. For example, Ryf et al. \cite{Ryf_JLightwaveTech2012} describe a mode-multiplexing interferometer incorporating simple phase masks that achieves $> 28$ dB mode selectivity. Instead of phase masks, spatial light modulators could also be used for high-fidelity mode conversion \cite{vanPutten_ApplOpt2008}. For another example, the lossless 4-pulse stacker used by Kienel et al. \cite{Kienel_OptLett2014} can easily be modified to produce 2 or 3 collinear pulses and incorporate double-pass AOMs and phase masks in the interferometer arms for frequency shifting and mode-conversion. Another possibility for providing delay and $f_{\textss{CEO}}$ shift with one device is to use a dazzler, as has recently been demonstrated for dual comb spectroscopy \cite{Znakovskaya_OptLett2014}. 

Notably absent from this list is the pulse shaper, now commonly employed by many 2D spectroscopists \cite{Grumstrup_OptExp2007,Shim_PCCP2009}. This is because the modulation frequencies must correspond to the frequency spacings of the higher-order modes, which for reasonable cavity lengths will be in the MHz regime, exceeding the update rate of both AOM-based and LCD-based pulse shapers.

\section{Conclusions}\label{sec:conclusions}

In conclusion, we have described methods to perform passively amplified 2D spectroscopy experiments using a frequency comb laser and optical cavities. As we have demonstrated for transient absorption spectroscopy in both isolated molecules \cite{Reber_Optica2016} and clusters \cite{Li_arXiv2016}, a large sensitivity improvement of several orders of magnitude is expected, enabling 2D spectroscopy in dilute molecular beams. Additionally, in the one cavity scheme, the pump-probe spatial overlap factors (equation (\ref{eqn:modematch})) relating the absolute size of the signal to strength of the molecule's nonlinear polarization can be known precisely, enabling greater quantification of 2D spectroscopy signals for fundamental studies or analytical chemistry applications. As in CE-TAS, the techniques are generally applicable to the UV, visible, and infrared regimes - wherever frequency combs with reasonable power can be generated and high reflectivity, low loss, low GDD mirrors can be fabricated. The necessary optical components and light sources have been demonstrated in all of these spectral ranges for other purposes, and the simultaneous coupling of a frequency comb to multiple higher-order modes has been previously used in the context of intracavity high-order harmonic generation \cite{Pupeza_PRL2014, Weitenberg_OptExp2011}.

The largest constraints on CE-2D schemes likely come from the simultaneous bandwidth that can be resonantly enhanced. This is not due to the attainable bandwidth of frequency comb sources or the reflectivity bandwidth of high-reflectivity, low loss cavity mirrors. Instead, it is set by the dispersion of the cavity mirrors. For a cavity with a finesse of 1000, to enhance 5 THz of bandwidth the net cavity group delay dispersion (GDD) must be controlled to better than 50 fs$^2$. This is quite feasible, but  resonantly enhancing 100's of THz of bandwidth for 2D spectroscopy experiments in the visible simultaneously is likely not feasible and 2D-electronic spectra will likely have to be acquired piecewise, scanning both the pump and probe. However, for many experiments in the IR, 5-10 THz (170-330 cm$^{-1}$) of bandwidth is more than sufficient, and recently there have been major breakthroughs in mid-IR cavity mirror technology \cite{Cole_Optica2016}.

The ability to perform 2D spectroscopy on cold gas-phase molecules allows the ultrafast spectroscopist to either (1) study a molecule of interest in a cold, collision free environment, recording the same optical signal as recorded in solution, or (2) study the dynamics of designer molecules that can only be made in cold supersonic expansions. We expect these new capabilities to allow for the study of the dynamics of molecules with unprecedented detail and control. Problems of interest include intramolecular vibrational relaxation \cite{Nesbitt_JPhysChem1996, Fang_JChemPhys2016}, many-body couplings and dynamics  of hydrogen bond networks \cite{Keutsch_PNAS2011}, solvation effects on the dynamics of small molecules \cite{Kungwan_PCCP2012}, exciton dynamics \cite{Sanders_Angewandte2016}, and nonadiabatic dynamics \cite{Polli_Nature2010, Allison_JChemPhys2012}. 

These CE-2D schemes can also be adapted to study condensed phase systems, as has been done for cavity-enhanced linear spectroscopy \cite{Vallance_bookchapter2014}. This would enable experiments to be performed on low-concentration solutions, sub monolayer films on surfaces \cite{Zanni_PNASCommentary2016, Kraack_JPCC2016}, or low temperature systems that must be excited very weakly \cite{Liu_PRB2011}. For example, including a prism into the cavity \cite{Pipino_ChemPhysLett1997} could allow for cavity-enhanced 2D attenuated total reflectance spectroscopy from thin layers on surfaces \cite{Kraack_JPCL2014}.

\section{Acknowledgements}

This work was supported by the National Science Foundation under award number 1404296 and the Air Force Office of Scientific Research under award number FA9550-16-1-0164. We thank T. C. Weinacht for useful discussions.

\vspace{0.5 cm}


\bibliography{/Users/tka3/Documents/bibliographies/masterbib.bib}
\bibliographystyle{unsrt}

\end{document}